\documentclass[aps,prd,preprint,twoside,tightenlines,showpacs,amsmath,amssymb,nofootinbib,floatfix,superscriptaddress]{revtex4}
\usepackage{graphicx}
\usepackage{amsfonts}
\usepackage{color}
\usepackage[colorlinks=true,urlcolor=blue,linkcolor=blue,citecolor=blue]{hyperref}
\usepackage[T1]{fontenc}
\usepackage{lmodern}

\begin{document}
\renewcommand{\theenumi}{(\alph{enumi})}

\title{\boldmath $\bar B^0$, $B^-$ and $\bar B^0_s$ decays into $J/\psi$ and $K \bar K$ or $\pi \eta$ }

\author{Wei-Hong Liang}
\email{liangwh@gxnu.edu.cn}
\affiliation{Department of Physics, Guangxi Normal University,
Guilin 541004, China}

\author{Ju-Jun Xie} \email{xiejujun@impcas.ac.cn}

\affiliation{Institute of Modern Physics, Chinese Academy of
Sciences, Lanzhou 730000, China}
\affiliation{State Key Laboratory of Theoretical Physics, Institute of Theoretical Physics, Chinese Academy of Sciences, Beijing 100190, China}

\author{E.~Oset}\email{oset@ific.uv.es}
\affiliation{Departamento de F\'{\i}sica Te\'orica and IFIC, Centro Mixto Universidad de Valencia-CSIC Institutos de Investigaci\'on de Paterna, Aptdo.
22085, 46071 Valencia, Spain}
\affiliation{Institute of Modern Physics, Chinese Academy of Sciences, Lanzhou 730000, China}

\date{\today}

\begin{abstract}
We study the  $\bar B^0_s \to J/\psi K^+ K^-$, $\bar B^0 \to J/\psi K^+ K^-$, $B^- \to J/\psi K^0 K^-$,
$\bar B^0 \to J/\psi \pi^0 \eta$ and $B^- \to J/\psi \pi^- \eta$ decays and compare their mass distributions with those obtained for the $\bar B^0_s \to J/\psi \pi^+ \pi^-$ and $\bar B^0 \to J/\psi \pi^+ \pi^-$. The approach followed consist in a factorization of the weak part and the hadronization part into a factor which is common to all the processes. Then what makes the reactions different are some trivial Cabibbo-Kobayashi-Maskawa matrix elements and the weight by which the different pairs of mesons appear in a primary step plus their final state interaction. These elements are part of the theory and thus, up to a global normalization factor, all the invariant mass distributions are predicted with no free parameters. Comparison is made with the limited experimental information available. Further comparison of these results with coming LHCb measurements will be very valuable to make progress in our understanding of the meson-meson interaction and the nature of the low lying scalar meson resonances, $f_0(500), f_0(980)$ and $a_0(980)$.
\end{abstract}

\maketitle

\section{Introduction}

Recent findings by the LHCb \cite{Aaij:2011fx,LHCb:2012ae,Aaij:2014emv}, Belle \cite{Li:2011pg}, CDF \cite{Aaltonen:2011nk}, and D0 \cite{Abazov:2011hv} collaboration, have shown that the $f_0(980)$ is produced in the $\bar B^0_s$ decays into $J/\psi$ and $\pi^+ \pi^-$ and no trace of the $f_0(500)$ is seen. At the same time the complementary reactions on the $\bar B^0$ decay into $J/\psi$ and $\pi^+ \pi^-$ \cite{Aaij:2013zpt,Aaij:2014siy}, have shown that a signal is seen for the $f_0(500)$ production and only a small fraction is observed for the $f_0(980)$. These findings have generated excitement in the hadron community. Studies of weak $B$ and $D$ decays into two final mesons had got some attention \cite{robert,bruno,cheng,bruno2,bruno3,lucio} in recent years. The order of magnitude of the $f_0(980)$ production was predicted in Ref. \cite{Colangelo:2010bg} using light cone QCD sum rules under the factorization assumption. After the experiments  quoted above, some studies were done based on the $q \bar q$ or tetraquark structure of the scalar mesons in Ref. \cite{Stone:2013eaa}. The $\bar B^0$ decay into $J/\psi$ and $\pi^+ \pi^-$
branching ratio was calculated in the QCD-improved factorization approach in Ref. \cite{sayahi},
where a two-meson distribution amplitude for the pion pair and final state interaction were considered. The chiral unitary approach was used in Ref. \cite{liang}, where predictions with no parameters fitted to the data were done for ratios of invariant mass distributions and the basic features of these experiments were well reproduced. Other work has also been done using the perturbative QCD approach in Ref. \cite{weiwang}.  More recently another approach has been used in Ref. \cite{hanhart} using effective Hamiltonians, transversity form factors and implementing the final sate interaction of the pions via the Omnes representation. In this latter paper the $\bar B^0_s$ decay into $J/\psi$ and $K^+ K^-$ is also studied and compared to experiment. The method uses experimental phase shifts for the Omnes representation and a few parameters fitted to the data. On the other hand the
$\bar B^0$ decay into $J/\psi$ and $K^+ K^-$ is not addressed there since
"this has an isovector component (corresponding e.g. to the $a_0(980)$ resonance) and  would
have to be described by a coupled-channel treatment of the $\pi \eta$  and $K^+ K^-$ $S$-waves."
Some work along these lines has been recently done in Ref. \cite{Albaladejo:2015aca}.
The chiral unitary approach for mesons \cite{npa,ramonet,kaiser,markushin,juanito,rios} is particularly suited for this purpose and has been used in many processes where the $f_0(500),f_0(980),a_0(980),\kappa(800)$ are produced, including some related to $D$ and $B$ weak decays \cite{ddec,bddec}. The purpose of this paper is to address the $K^+ K^-$ production in the $\bar B^0_s$ decays into $J/\psi$ and two kaons and the $K^+ K^-$ and $\pi \eta$ production in the $\bar B^0$ decay into $J/\psi$ and this pair of mesons.
Simultaneously, we shall also address the $B^- \to J/\psi K^0 K^-$ and $B^- \to J/\psi \pi^- \eta$ decays.

Experimentally, we have information in Refs. \cite{Aaij:2013mtm,Aaij:2013orb}.
In Ref. \cite{Aaij:2013mtm} the $\bar B^0$ decay into $J/\psi$ and $K^+ K^-$ is
studied. Rates for this branching ratio are provided  and also for the
$\bar B^0 \to J/\psi a_0(980), a_0(980) \to K^+ K^-$. However, the $\pi^0
\eta$ decay channel of the  $a_0(980)$ is not investigated.
In Ref. \cite{Aaij:2013orb} a full phase space partial wave analysis of the $\bar B_s^0 \to J/\psi K \bar K$ is done, including $S, P$ and $D$ waves.
The  $\bar B^0_s$ decay into $J/\psi$ and $f'_2(1525)$ with
$f'_2(1525)$ decaying into  $K^+ K^-$ is addressed there \cite{Aaij:2013orb}. In this case, the
$K^+ K^-$ is in $D$-wave and in the present paper we only consider $S$-wave.
The work of Ref. \cite{Aaij:2013orb} on $\bar B^0_s \to J/\psi K^+ K^-$ separates the $S$-wave $K^+ K^-$ below the $\phi$
region and this allows for a comparison with our predictions. Note that the results of Refs. \cite{Aaij:2013mtm,Aaij:2013orb}
are based on $1/fb$ of $pp$ collision data accumulated by LHCb during 2011. Further improvement can be anticipated from studies including the
$2/fb$ of data accumulated during 2012 and with new data now being taken in the LHC Run 2.

A good thing of the use of the
chiral unitary approach for these reactions is that one can obtain mass
distributions up to an arbitrary normalization and relate the different
distributions with no parameters fitted to the data. This is due to the
unified picture that the chiral unitary approach provides for the
scattering of mesons. In this sense, predictions on the coming
measurements should be most welcome, and if supported by experiment, it
can give us elements to dig into the nature of the low lying scalar
mesons, which in this picture emerge as dynamically generated from the
interaction of pseudoscalar mesons using a meson-meson interaction
derived from the chiral Lagrangians \cite{Gasser:1983yg,Bernard:1995dp}.

\section{formalism}

\begin{figure}[b!]\centering
\includegraphics[height=3.0cm,keepaspectratio]{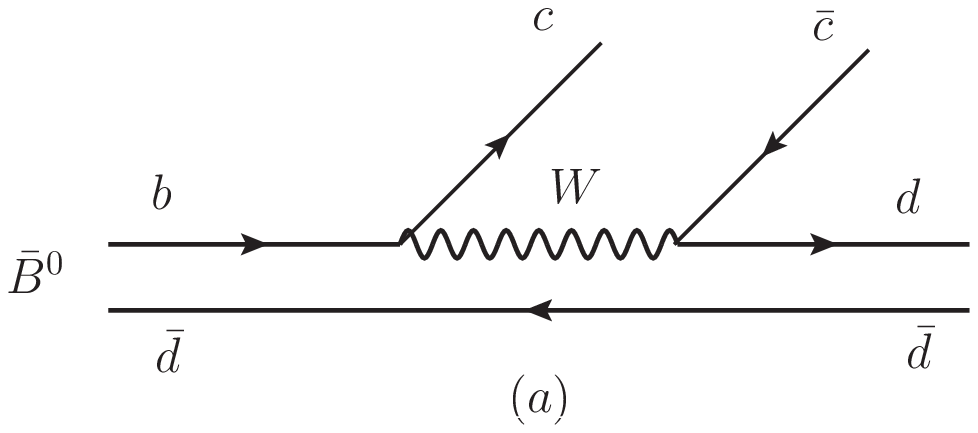}
\includegraphics[height=3.2cm,keepaspectratio]{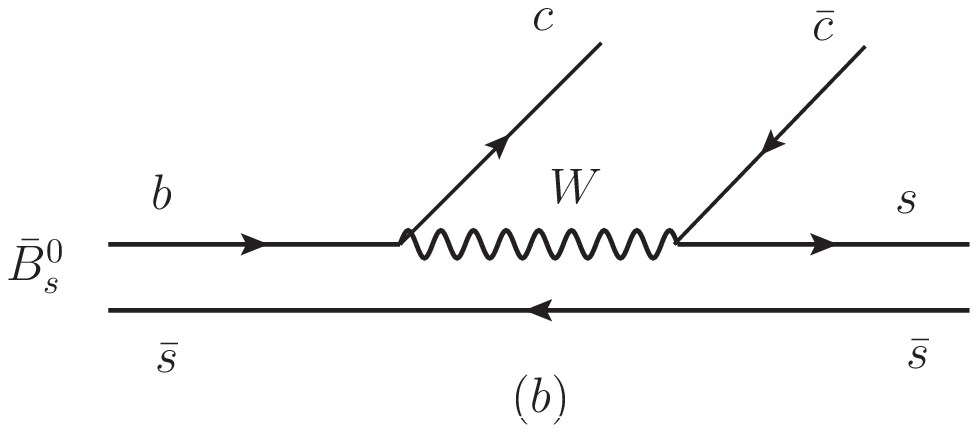}
\caption{Diagrams for the decay of $\bar B^0$ and $\bar B^0_s$ into $J/\psi$ and a primary $q\bar q$ pair, $d\bar d$ for $\bar B^0$ and $s\bar s$ for $\bar B^0_s$.
\label{fig:fig1}}
\end{figure}

Following Refs. \cite{Stone:2013eaa,liang}, in Fig. \ref{fig:fig1} we draw the diagrams at the quark level that are responsible for the $\bar B^0$ and $\bar B^0_s$ decays into $J/\psi$ and another pair of quarks: $d \bar d$ in the case of the $\bar B^0$ decay and $s \bar s$ in the case of $\bar B^0_s$ decay. The first process involves the $V_{cd}$, Cabibbo suppressed Cabibbo-Kobayashi-Maskawa matrix element, and the second one the  $V_{cs}$ Cabibbo allowed one, which makes the widths large in the second case compared to the first one.

\begin{figure}[b!]\centering
\includegraphics[height=2.1cm,keepaspectratio]{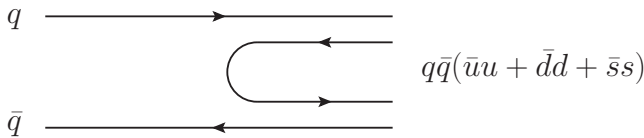}
\caption{Schematic representation of the hadronization $q\bar q \to q\bar q (\bar u u +\bar d d +\bar s s)$.\label{fig:fig2}}
\end{figure}

 Following the work of Ref. \cite{liang} we put together in a factor $V_P$ all elements which are common in the two decays. Next, in order to produce two mesons the final $q \bar q$ state has to hadronize, which we implement adding a  $\bar q q $ pair with the quantum numbers of the vacuum
$\bar u u+ \bar d d+ \bar s s$, see Fig. \ref{fig:fig2}. Then we define the matrix $M$ for the  $q \bar q$ elements

\begin{equation}
M=\left(
           \begin{array}{ccc}
             u\bar u & u \bar d & u\bar s \\
             d\bar u & d\bar d & d\bar s \\
             s\bar u & s\bar d & s\bar s \\
           \end{array}
         \right),
\end{equation}
which has the property
\begin{equation}\label{eq:2}
M\cdot M=M \times (\bar u u +\bar d d +\bar s s).
\end{equation}

Next we write the $ q \bar q$ matrix elements of $M$ in terms of the physical mesons and then we can replace the matrix $M$ by the matrix $\Phi$ given by
\begin{widetext}
\begin{equation}\label{eq:phimatrix}
\Phi = \left(
           \begin{array}{ccc}
             \frac{1}{\sqrt{2}}\pi^0 + \frac{1}{\sqrt{3}}\eta + \frac{1}{\sqrt{6}}\eta' & \pi^+ & K^+ \\
             \pi^- & -\frac{1}{\sqrt{2}}\pi^0 + \frac{1}{\sqrt{3}}\eta + \frac{1}{\sqrt{6}}\eta' & K^0 \\
            K^- & \bar{K}^0 & -\frac{1}{\sqrt{3}}\eta + \sqrt{\frac{2}{3}}\eta' \\
           \end{array}
         \right),
\end{equation}
\end{widetext}
where we are taking the standard mixing of the $\eta$ and  $\eta'$ in terms of a singlet and an octet of SU(3),
\begin{equation}\label{eq:eta-etapmixing}
 \eta  =  \frac{1}{3} \eta_1 + \frac{2\sqrt{2}}{3} \eta_8, ,~~~~~~\eta'  =  \frac{2\sqrt{2}}{3} \eta_1 -  \frac{1}{3} \eta_8.
\end{equation}

\begin{figure}[tb]\centering
\includegraphics[scale=0.48]{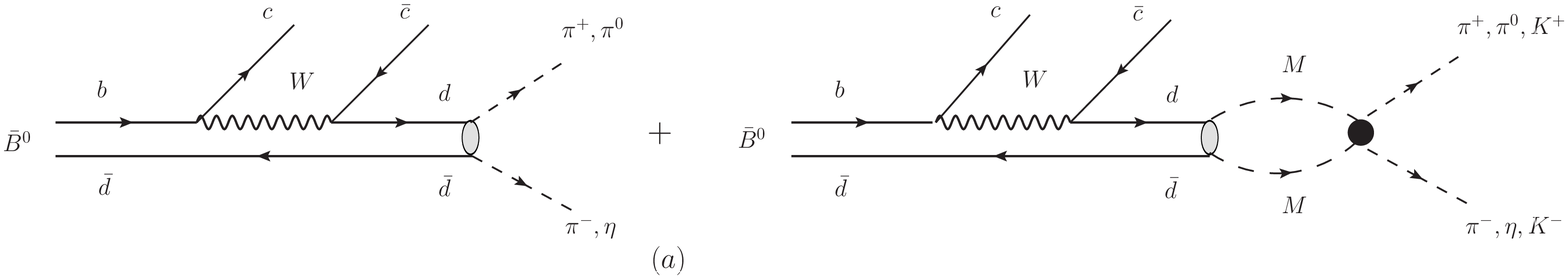}
\includegraphics[scale=0.48]{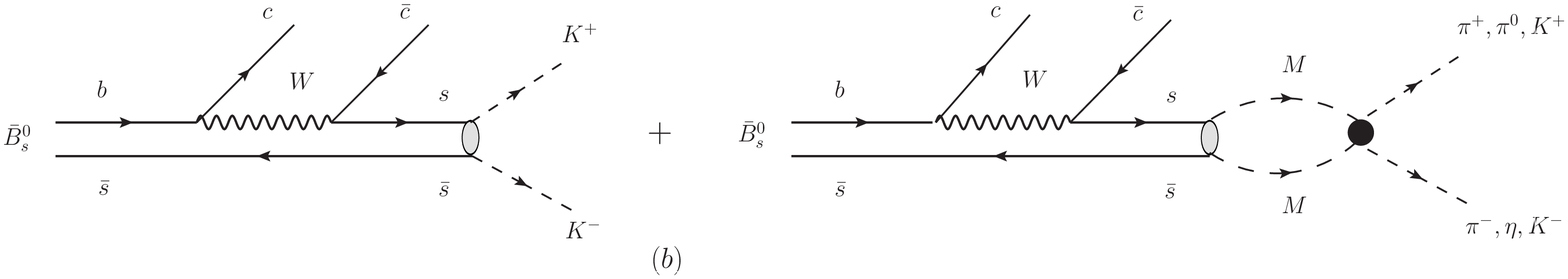}
\caption{Diagrammatic representations of the production of $\pi^+ \pi^-$, $\pi^0 \eta$ and $K^+ K^-$ via direct plus rescattering mechanisms in $\bar B^0$ (a) and $\bar B ^0_s$ (b) decays.
\label{fig:pipi_production}}
\end{figure}

The hadronization leads us to\footnote{There are small changes here with respect to Ref. \cite{liang}, where the singlet in the $\Phi$ matrix was ignored.}
\begin{align}\label{eq:PhiPhi_elements}
d\bar d (\bar u u +\bar d d +\bar s s) & \equiv  \left( \phi \cdot \phi \right)_{22}=\pi^- \pi^+ +\frac{1}{2}\pi^0 \pi^0
+\frac{1}{3}\eta \eta -\frac{2}{\sqrt{6}}\pi^0 \eta +\bar K^0 K^0 ,\nonumber \\
s\bar s (\bar u u + \bar d d +\bar s s) & \equiv  \left( \phi \cdot \phi \right)_{33}=K^- K^+ + \bar K^0 K^0  +\frac{1}{3}\eta \eta .
\end{align}
where the $\eta'$ terms have been neglected because the $\eta'$ is too massive and has not effect here.
The decomposition of the former formulas tell us the weight by which a pair of pions are produced in the first step.  The next step consists in taking into account the interaction of the mesons produced to finally obtain the desired couple of mesons. This is represented in Fig. \ref{fig:pipi_production}.

The amplitudes for a final production of the different meson pairs are  given by\footnote{Here there is an extra factor of two in the $\pi^0 \pi^0$ and $\eta \eta$ terms with respect to Ref. \cite{liang} because of the two possibilities to create the two identical particles with the $\pi^0 \pi^0$ and $\eta \eta$ operators of Eq. (\ref{eq:PhiPhi_elements}). The changes in the numerical results for $\bar B^0$ decay are of the order of 15\% and there are no changes for the $\bar B^0_s$ decay. }
\begin{align}\label{eq:t_B2Jpsipipi}
t(\bar B^0 \to J/\psi \pi^+ \pi^-) & = V_P V_{cd}
\left(
1+G_{\pi^+ \pi^-} t_{\pi^+ \pi^- \to \pi^+ \pi^-} +2 \frac{1}{2} \frac{1}{2} G_{\pi^0 \pi^0} t_{\pi^0 \pi^0 \to \pi^+ \pi^-}  \right.\nonumber \\
&~~~~~~~~~~~~~~ \left. +G_{K^0 \bar K^0} t_{K^0 \bar K^0 \to \pi^+ \pi^-} +2 \frac{1}{3} \frac{1}{2} G_{\eta \eta} t_{\eta \eta \to \pi^+ \pi^-}\right),
\end{align}

\begin{equation}\label{eq:t_B2Jpsipieta}
 t(\bar B^0 \to J/\psi \pi^0 \eta)  = V_P V_{cd}
\left(-\frac{2}{\sqrt{6}}-\frac{2}{\sqrt{6}} G_{\pi^0 \eta} t_{\pi^0 \eta \to \pi^0 \eta} +G_{K^0 \bar K^0} t_{K^0 \bar K^0\to\pi^0 \eta}\right),
\end{equation}

\begin{align}\label{eq:t_B2JpsiKK}
t(\bar B^0 \to J/\psi K^+ K^-) & = V_P V_{cd}
\left(
G_{\pi^+ \pi^-} t_{\pi^+ \pi^- \to K^+ K^-} +2 \frac{1}{2} \frac{1}{2} G_{\pi^0 \pi^0} t_{\pi^0 \pi^0 \to K^+ K^-}
  \right.\nonumber \\
& \left. +2 \frac{1}{3} \frac{1}{2} G_{\eta \eta} t_{\eta \eta \to K^+ K^-} -\frac{2}{\sqrt{6}} G_{\pi^0 \eta} t_{\pi^0 \eta \to K^+ K^-} +G_{K^0 \bar K^0} t_{K^0 \bar K^0\to K^+ K^-}
\right),
\end{align}

\begin{equation}\label{eq:t_Bs2Jpsipipi}
t(\bar B^0_s \to J/\psi \pi^+ \pi^-)  = V_P V_{cs} \left( G_{K^+ K^-} t_{K^+ K^- \to \pi^+ \pi^-} +  G_{K^0 \bar K^0} t_{K^0 \bar K^0 \to \pi^+ \pi^-}  +2 \frac{1}{3} \frac{1}{2} G_{\eta \eta} t_{\eta \eta \to \pi^+ \pi^-} \right),
\end{equation}
\begin{align}\label{eq:t_Bs2JpsiKK}
t(\bar B^0_s \to J/\psi K^+ K^-) & = V_P V_{cs} \left(1+ G_{K^+ K^-} t_{K^+ K^- \to K^+ K^-} +  G_{K^0 \bar K^0} t_{K^0 \bar K^0 \to K^+ K^-}  \right.\nonumber \\
&~~~~~~~~~~~~ \left. +2 \frac{1}{3} \frac{1}{2} G_{\eta \eta} t_{\eta \eta \to K^+ K^-} \right),
\end{align}
where $G_i$ are the loop functions of two meson propagators. In Eqs. (\ref{eq:t_B2Jpsipipi})-(\ref{eq:t_Bs2JpsiKK}) we have taken into account that for the identical particles one has a factor of two from the two ways to match the two identical particles of the operator in Eqs. (\ref{eq:PhiPhi_elements}) with the two mesons produced and a factor $1/2$ in the $G$ function. The $t_{ij}$ are the scattering matrices and they are calculated in Ref. \cite{liang} following Ref. \cite{npa}.

In the case of $\bar B^0 \to J/\psi K^+ K^-$ decay, we have contributions from $I=0$ and $I=1$. It is interesting to split the contributions. This is easily done taking all the $G_{\pi \pi}$ equal, and also $G_{K^0 K^0} = G_{K^+ K^-}$ and rewriting $t_{K^0 \bar K^0\to K^+ K^-}$ as
\begin{equation}\label{eq:t_KK}
t_{K^0 \bar K^0\to K^+ K^-} = \frac{1}{2} t_{K^0 \bar K^0\to K^+ K^-} + \frac{1}{2} t_{K^+  K^-\to K^+ K^-}
 + \frac{1}{2} t_{K^0 \bar K^0\to K^+ K^-} - \frac{1}{2} t_{K^+  K^-\to K^+ K^-},
\end{equation}
where the first two terms are in $I=0$ while the last two terms are in $I=1$. Then we split Eq. (\ref{eq:t_B2JpsiKK}) into
\begin{equation}\label{eq:t_B2JpsiKKI01}
t(\bar B^0 \to J/\psi K^+ K^-)= t^{I=0} + t^{I=1}
\end{equation}
with
\begin{align}\label{eq:t_B2JpsiKKI0}
t^{I=0} & = V_P V_{cd}
\left[
G_{\pi \pi} t_{\pi^+ \pi^- \to K^+ K^-} + \frac{1}{2} G_{\pi \pi} t_{\pi^0 \pi^0 \to K^+ K^-}
+  \frac{1}{3} G_{\eta \eta} t_{\eta \eta \to K^+ K^-}
\right.\nonumber \\
&~~~ \left. +~G_{K \bar K} \left( \frac{1}{2} t_{K^0 \bar K^0\to K^+ K^-} + \frac{1}{2} t_{K^+ K^-\to K^+ K^-} \right)\right],
\end{align}
\begin{align}\label{eq:t_B2JpsiKKI1}
t^{I=1} & = V_P V_{cd}
\left[
-\frac{2}{\sqrt{6}} G_{\pi^0 \eta} t_{\pi^0 \eta \to K^+ K^-}
+G_{K \bar K} \left( \frac{1}{2} t_{K^0 \bar K^0\to K^+ K^-} - \frac{1}{2} t_{K^+ K^-\to K^+ K^-} \right)\right].
\end{align}

\begin{figure}[b!]\centering
\includegraphics[height=3.0cm,keepaspectratio]{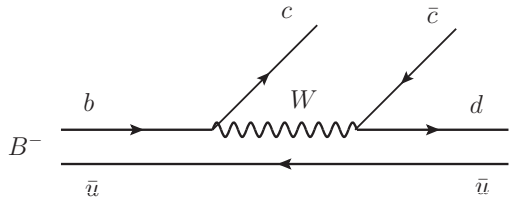}
\caption{Diagrams for the decay of $B^-$ into $J/\psi$ and a primary $d\bar u$ pair.
\label{fig:figB-decay-QuarkLevel}}
\end{figure}

The $B^-$ decays proceed in an analogous way. The basic quark diagram is now given in Fig. \ref{fig:figB-decay-QuarkLevel}.
The hadronization leads us now, neglecting $\eta'$, to
\begin{equation}\label{eq:PhiPhi_elements_21}
d\bar u (\bar u u +\bar d d +\bar s s) \equiv  \left( \phi \cdot \phi \right)_{21}=\frac{2}{\sqrt{3}}\pi^- \eta + K^0 K^- ,
\end{equation}
from where, taking into account the final state interaction of the mesons, we obtain
\begin{equation}\label{eq:t_B-2Jpsipieta}
t(B^- \to J/\psi \pi^- \eta) = V_P V_{cd}
\left(
\frac{2}{\sqrt{3}}+\frac{2}{\sqrt{3}} G_{\pi^- \eta} t_{\pi^- \eta \to \pi^- \eta} +G_{K^0 K^-} t_{K^0 K^- \to \pi^- \eta} \right),
\end{equation}
\begin{equation}\label{eq:t_B-2JpsiK0K-}
t(B^- \to J/\psi K^0 K^-) = V_P V_{cd}
\left(1+G_{K^0 K^-} t_{K^0 K^- \to K^0 K^-}+
\frac{2}{\sqrt{3}} G_{\pi^- \eta} t_{\pi^- \eta \to K^0 K^-} \right).
\end{equation}

There is no need to recalculate the meson-meson scattering matrix $t_{ij}$, since using isospin symmetry and the fact
that $|K^0 K^-\rangle = -| 1,1\rangle $ of isospin, we have
\begin{align}
t_{\pi^- \eta \to \pi^- \eta} &= t_{\pi^0 \eta \to \pi^0 \eta} \equiv t_{\pi \eta \to \pi \eta}^{I=1}, \label{eq:t_pieta-pieta}\\
t_{K^0 K^- \to \pi^- \eta} &= -~t_{K\bar K \to \pi \eta}^{I=1} = -\sqrt{2}~ t_{K^0 \bar K^0 \to \pi^0 \eta}, \label{eq:t_K0K-pieta}\\
t_{K^0 K^- \to K^0 K^-} &= t_{K\bar K \to K\bar K}^{I=1} = t_{K^+ K^- \to K^+ K^-}- t_{K^+ K^- \to K^0 \bar K^0}.\label{eq:t_K0K-K0K}
\end{align}
In particular one can see that the $t(B^- \to J/\psi \pi^- \eta)$ amplitude of Eq. (\ref{eq:t_B-2Jpsipieta}) is $- \sqrt{2}$ times the amplitude for
$t(\bar B^0 \to J/\psi \pi^0 \eta)$ and hence its rate of production will be a factor of two bigger.

One final observation is the fact that in a $0^- \to 1^- 0^+$ transition we shall need a $L'=1$ for the $J/\psi$ to match angular momentum conservation. Hence, $V_P = A~p_{J/\psi} \cos \theta$, and we assume $A$ to be constant (equal $1$ in the calculations). Thus,
\begin{equation}\label{eq:dGamma}
  \frac{d \Gamma}{d M_{\rm inv}}=\frac{1}{(2\pi)^3}\frac{1}{4M_{\bar B_j}^2}\frac{1}{3}p_{J/\psi}^2 p_{J/\psi} \tilde{p}_{\pi} {\overline{ \sum}} \sum \left| \tilde{t} (\bar B^0_j \to J/\psi \pi^+ \pi^-) \right|^2,
\end{equation}
where the factor $1/3$ is coming from the integral of $\cos^2 \theta$ and $\tilde{t} (\bar B^0_j \to J/\psi \pi^+ \pi^-)$ is $t (\bar B^0_j \to J/\psi \pi^+ \pi^-)/(p_{J/\psi} \cos \theta)$, which depends on the $\pi^+ \pi^-$ invariant mass. In
Eq.~(\ref{eq:dGamma}) $p_{J/\psi}$ is the $J/\psi$ momentum in the global CM frame ($\bar B$ at rest) and $\tilde{p}_{\pi}$ is the pion momentum in the $\pi^+ \pi^-$ rest frame.

\section{Results}

\begin{figure}[thb]\centering
\includegraphics[scale=0.35]{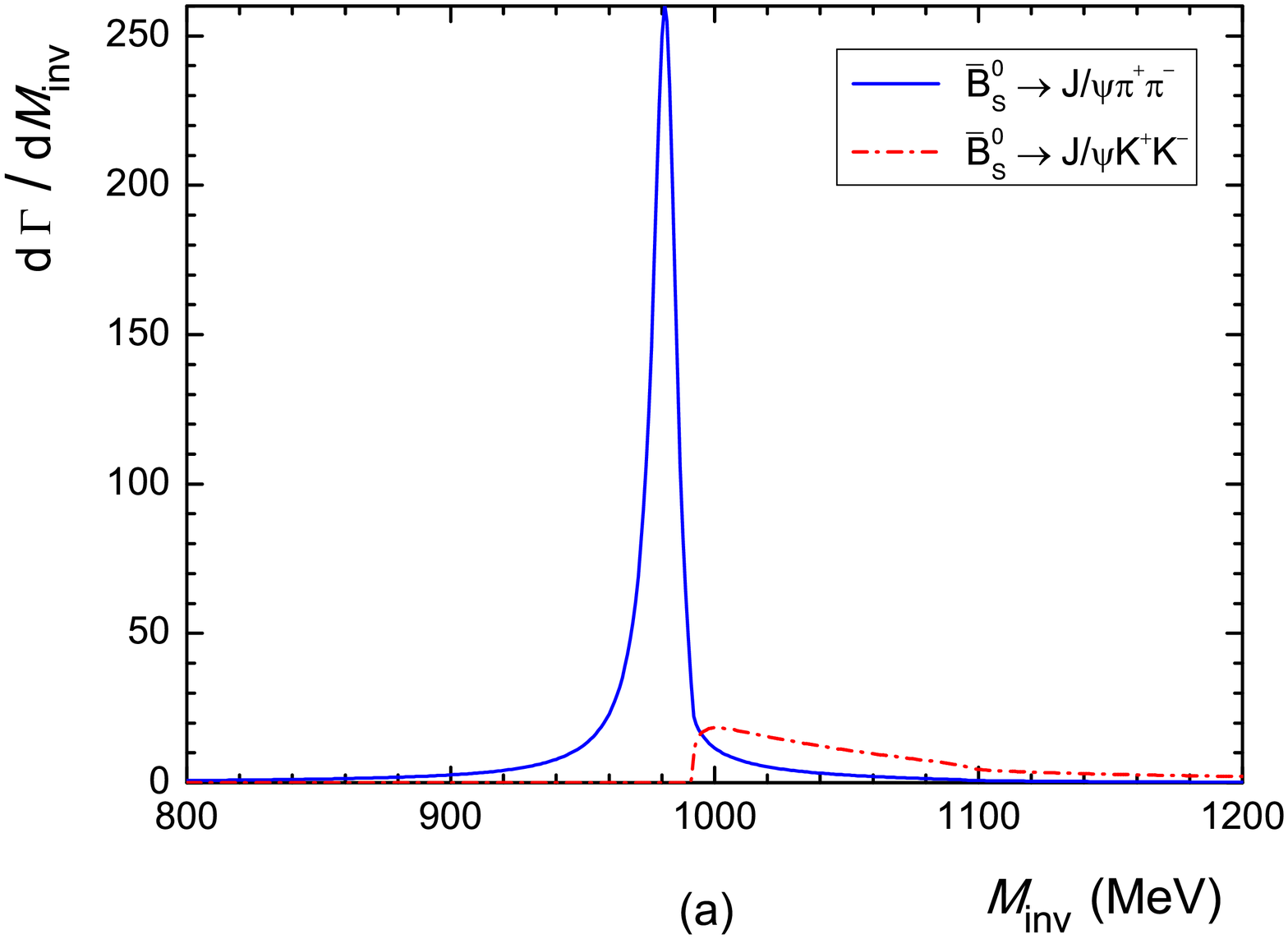}
\includegraphics[scale=0.35]{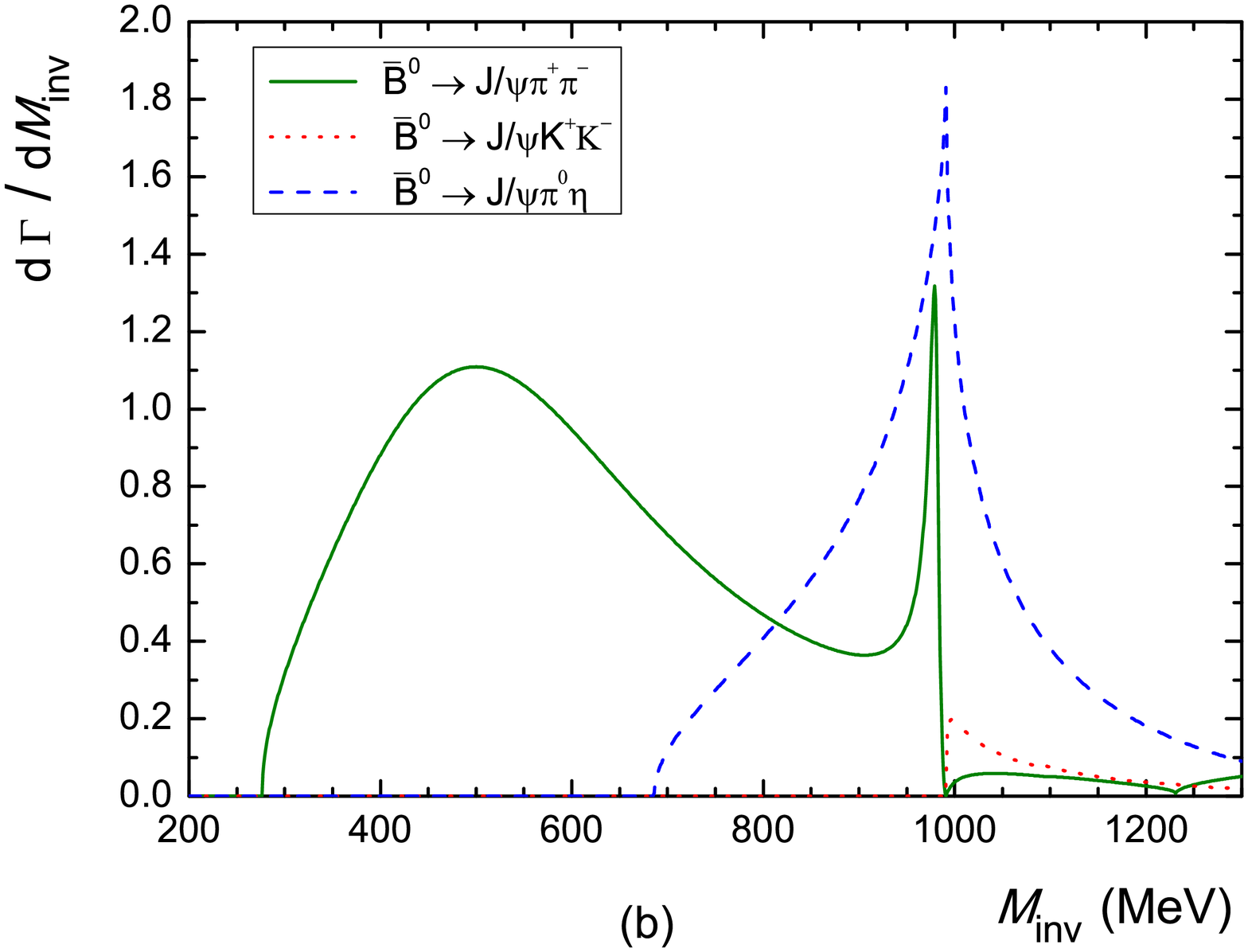}
\caption{(a) $\pi^+ \pi^-$, $K^+ K^-$ invariant mass distributions for the $\bar B_s^0 \to J/\psi \pi^+ \pi^-$, $J/\psi K^+ K^-$ decays;
(b) $\pi^+ \pi^-$, $\pi^0 \eta$, $K^+ K^-$ invariant mass distributions for the $\bar B^0 \to J/\psi \pi^+ \pi^-$, $J/\psi K^+ K^-$, $J/\psi\pi^0 \eta$ decays.
\label{fig:mass_distribution}}
\end{figure}

\begin{figure}[thb]\centering
\includegraphics[scale=0.35]{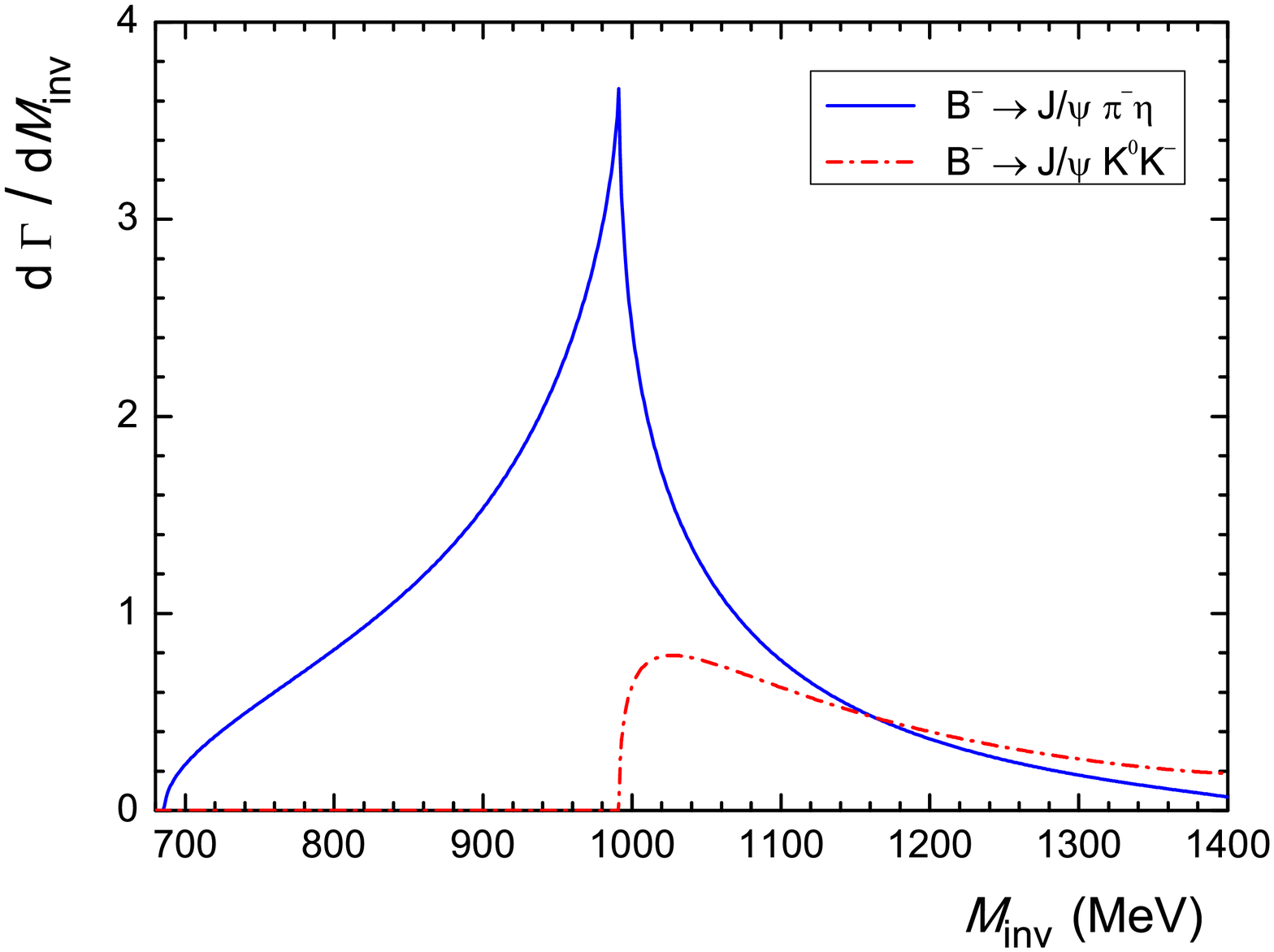}
\caption{$\pi^- \eta$ and $K^0 K^-$ invariant mass distributions for the $B^- \to J/\psi \pi^- \eta$ and $B^- \to J/\psi K^0 K^-$ decays.
\label{fig:mass_distribution_pi2eta}}
\end{figure}

Our results are summarized in Figs. \ref{fig:mass_distribution} (a) and (b), and Fig. \ref{fig:mass_distribution_pi2eta}. In Fig. \ref{fig:mass_distribution} (a) we show the
results for the $\bar B^0_s \to J/\psi \pi^+ \pi^-$ together with those for
$\bar B^0_s \to J/\psi K^+ K^-$. The former results are those calculated  in
Ref. \cite{liang}. The latter are new. As we can see, the $K^+ K^-$ distribution gets
maximum strength close to the  $K^+ K^-$ threshold and then falls down
gradually. This is due to the effect of the $f_0(980)$ resonance below
threshold. In this case we started from an $s \bar s$ quark state, which
has isospin zero, and the strong interaction hadronization conserves it.
So, even if $K^+ K^-$ could be in $I=0,1$, the process of formation
guarantees that this is an $I=0$ state and the shape of the distribution is
due to the $f_0(980)$. The strength is small compared to the one of the
$f_0(980)$ at its peak, but the integrated strength over the invariant
mass of $K^+ K^-$ is of the same order of magnitude as that for the
strength below the peak of the $f_0(980)$ going to $\pi^+
\pi^-$.\footnote{The apparent difference in the peak between the present
distribution and the results in Ref. \cite{liang} are due to the fact that
the distribution was folded with the experimental resolution of 20 MeV
in Ref. \cite{liang}.} However, we should take into account that we are
calculating only the $S$-wave part of the  $K^+ K^-$ spectrum, hence,
contributions from $\phi$ ($P$-wave), $f'_2(1525)$ ($D$-wave) etc, are
not evaluated. It is interesting to compare this with experiment. First
by integrating the strength of the $K^+ K^-$ distribution over its
invariant mass we find a ratio
\begin{equation}\label{eq:ratio_BsKK}
 \frac{\mathcal{B} [ \bar B^0_s \to J/\psi K^+ K^-]}{\mathcal{B} [\bar B^0_s \to J/\psi f_0(980); f_0(980)
\to \pi^+ \pi^- ]} = 0.34  \pm 0.03,
\end{equation}
where we have added an estimated 10 \% theoretical uncertainty.

By taking into account that the rates for $f_0(980)$ and $\phi$ are
\begin{align}\label{eq:ratio_Bstof0}
& \mathcal{B} [\bar B^0_s \to J/\psi f_0(980); f_0(980) \to \pi^+ \pi^-] =  (1.39 \pm 0.14  )\times 10^{-4},\nonumber \\
& \mathcal{B} [\bar B^0_s \to J/\psi \phi] =  (1.07 \pm 0.09) \times 10 ^{-3},
\end{align}
we find
\begin{equation}\label{eq:ratio_BsKKtoBsphi}
\frac{\mathcal{B}[\bar B^0_s \to J/\psi K^+ K^-]}{\mathcal{B} [\bar B^0_s \to J/\psi \phi]}=0.044 \pm 0.007 ,
\end{equation}
where we have summed the relative errors in quadrature.

If we stick to a band of energies $m_{\phi}\pm 12$ MeV we find
\begin{equation}\label{eq:ratio_BsKKtoBsphi-2}
\frac{\mathcal{B}[\bar B^0_s \to J/\psi K^+ K^-](S{\text{-wave}})}{\mathcal{B} [\bar B^0_s \to J/\psi \phi; \phi
\to K^+ K^- ]}= 0.017 \pm 0.003 ,
\end{equation}
where the branching fraction of $0.489$ for $\phi$ decay into $K^+ K^-$ has been
taken.  This value is in agreement with experiment \cite{Aaij:2013orb}, $(1.1 \pm 0.1 ^{+0.2}_{-0.1}) \times 10^{-2}$ within errors.
These experimental
numbers are consistent with, and improve, earlier measurements from CDF
\cite{Aaltonen:2012ie} and ATLAS \cite{Aad:2012kba}. They are also in
agreement with the results of Ref. \cite{hanhart} and the estimations prior
to the experiment in Ref. \cite{Stone:2008ak}.
On the other hand,  by looking at Fig. 17 of  Ref. \cite{Aaij:2013orb} one
can compare our results
for the  $K^+ K^-$ mass distribution with the contribution of the
$f_0(980)$ in that figure, and
the accumulated strength around threshold and the fall down with
increasing invariant mass agree fairly well, although our distribution
falls faster.

We come back now to the decays of the $\bar B^0$.  In Fig. \ref{fig:mass_distribution} (b) we show the
results for the
$\bar B^0$ into $J/\psi$ and $\pi^+ \pi^-$,  $J/\psi$ and  $K^+ K^-$ and
$J/\psi$ plus $\pi^0 \eta$.
In this case we had the hadronization of a $d \bar d$ state , which
contains $I=0,1$. The $\pi^+ \pi^-$ in $S$-wave, however, can only be in $I=0$,
hence the peaks for this distribution reflect again the $f_0(500)$ and $f_0(980)$
excitation. We should note that the normalization in Figs. \ref{fig:mass_distribution} (a) and (b) is the
same. Hence, the difference in size mostly reflects the differences
between the CKM matrix elements. We should note that because of the
experimental resolution this peak would not appear so narrow in the
experiments, but the integrated strength should be comparable, and this
was already done in Ref. \cite{liang}. Note that in any case, the strength of
the $f_0(980)$ excitation is very small compared to that of the
$f_0(500)$ (the broad peak to the left) as was already noted in the
experiments. The new results in this paper are for the  $K^+ K^-$ and
$\pi^0 \eta$ distributions. The $\pi^0 \eta$ distribution has a sizeable
strength, much bigger than that for the  $f_0(980)$ and reflects the
$a_0(980)$ excitation. This prediction is tied exclusively to the
weights of the starting meson meson channels in Eqs. (\ref{eq:PhiPhi_elements}) and the
scattering matrices appearing in  Eqs. (\ref{eq:t_B2Jpsipipi}). Hence, this is a
prediction of this approach, not tied to any experimental input.

In Fig. \ref{fig:mass_distribution}(b) we have plotted only the $S$-wave contribution. In the case of $\pi^0 \eta$, there is no relevant $P$-wave contribution in the $M_{\rm inv}$ region of the figure \cite{Abele:1999tf,Chung:2002fz}. However, this is not the case for $\pi^+ \pi^-$, which can couple to the $\rho$ meson in $P$-wave and give a large contribution. This is indeed the case and it was evaluated in Ref. \cite{Bayar:2014qha} (see Fig. 5 of that paper). The $\rho$ contribution peaks around 770 MeV, and has larger strength than the $f_0(500)$ contribution, but at invariant masses around 500 MeV and bellow, the strength of the $f_0(500)$ dominates the one of the $\rho$.

The $K^+ K^-$ distribution in the $\bar B^0$ decay is now both in $I=0$ and $I=1$, hence it
reflect the effects of both the $f_0(980)$ and the $a_0(980)$
resonances. In this case, if we draw a smooth curve below the $f_0(980)$
peak to separate it from the $f_0(500)$ contribution, we find,
\begin{equation}\label{eq:ratio_B2KKtof0}
\frac{\mathcal{B}[\bar B^0 \to J/\psi K^+ K^-]}{\mathcal{B} [\bar B^0 \to J/\psi f_0(980); f_0(980) \to
\pi^+ \pi^-]}=0.53 \pm 0.05 .
\end{equation}
The strength of this ratio is a bit larger now than the corresponding one for
$\bar B^0_s$ decay,  given in Eq. (\ref{eq:ratio_BsKK}).

In Ref. \cite{Aaij:2013mtm} we also find the branching ratio for
$\bar B^0 \to J/\psi a_0(980)$ with $a_0(980)$ decaying into $K^+ K^-$,
\begin{equation}\label{eq:Br_B2Jpsia0}
 \mathcal{B} [\bar B^0 \to J/\psi a_0(980), a_0(980) \to K^+ K^-] =  (4.7\pm 3.31\pm
0.72)\times 10^{-7}.
\end{equation}
It is unclear how this is obtained because, as discussed above, both the
$f_0(980)$ and
$a_0(980)$ resonances contribute to the $K^+ K^-$ distribution.

With the caveat about not comparing exactly the same magnitude we can
make an estimate of the $a_0(980)$ production rate by taking a background below the $\pi^0
\eta$ distribution as done in Ref. \cite{ddec} and using the former equations
we get a rate
\begin{equation}\label{eq:ratio_BsKKtoBsphi-2}
\mathcal{B}( \bar B^0  \to J/\psi \pi^0 \eta)= ( 2.2 \pm 0.2) \times 10^{-6}.
\end{equation}
If we multiply the latter value by the ratio $\Gamma(K^+ K^-)/ \Gamma(\eta \pi)
= \frac{1}{2}  (0.183 \pm 0.024)$ \cite{Agashe:2014kda}, we obtain $( 2.0 \pm 0.3) \times 10^{-7} $,
which is agreement with the value of Eq. (\ref{eq:Br_B2Jpsia0}) within errors.

We would also like to compare our $K^+ K^-$ mass distribution with the one of Fig.
13 of Ref. \cite{Aaij:2013mtm}. The experimental distribution after
subtraction of combinatorial background and misidentified events has in
principle large errors, which will be improved in coming experiments,
but we can see  that the shape of the distribution agrees  qualitatively
with ours, with a relatively faster fall down of the experimental one.

We come now to the results for $B^-$ decay which we plot in Fig. \ref{fig:mass_distribution_pi2eta}.
The scale of the figure
is the same as in Fig. \ref{fig:mass_distribution}.
As discussed in the former section, the strength for the $\pi^- \eta$ mass distribution in
$B^- \to J/\psi \pi^- \eta$ is twice as big as the one of $\bar B^0 \to J/\psi \pi^0 \eta$.
The strength of the $K^0 K^-$ mass distribution at the peak is however about four times bigger than the one for $K^+ K^-$ in the $\bar B^0$ decay.
We also observe that the position of the peak has moved to higher invariant masses compared to the $\bar B^0$ or $\bar B_s^0$ cases.
Both features find a natural explanation in the fact the the $K^0 K^-$ distribution in the $B^-$ decay
is due to the $a_0(980)$, which is seen in the figures, is much wider than that of the $f_0(980)$.
We should also note that the shape of the $K^+ K^-$ distribution
in the $\bar B^0$ case is also a bit different, sticking more towards the $K\bar K$ threshold.
We also see that the $f_0(980)$ distribution in this decay has
a different shape to that in the $\bar B_s^0$ decay, with zero strength around 1000 MeV.
It is clear that there are now interferences of the different terms contributing to the amplitude in Eq. (\ref{eq:t_B2JpsiKKI01}).

\begin{figure}[thb]\centering
\includegraphics[scale=0.35]{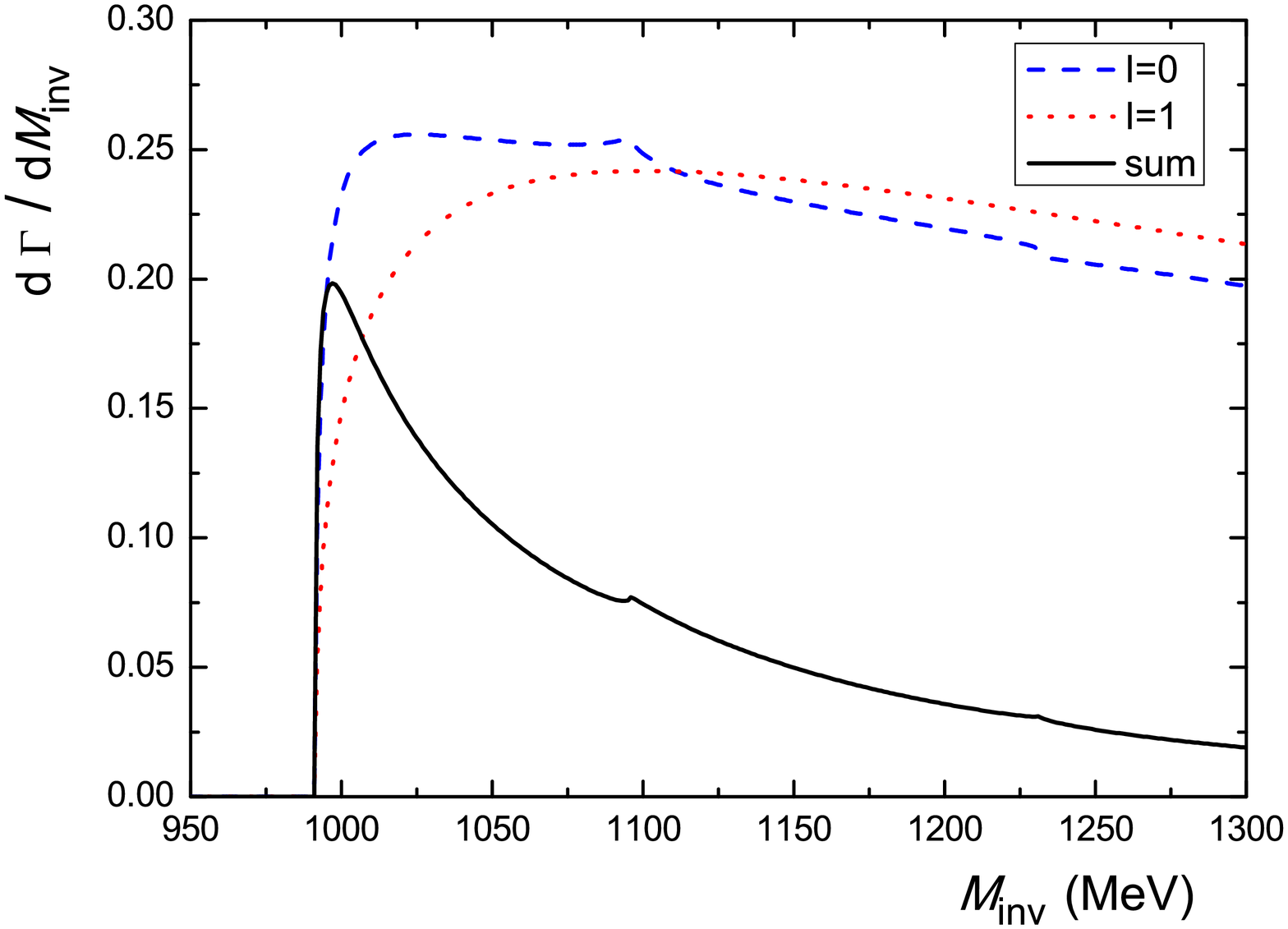}
\caption{$K^+ K^-$ invariant mass distributions for the $\bar B^0 \to J/\psi K^+ K^-$ decay. The $I=0$, $I=1$ individual contribution and the total, given by the sum of the amplitudes are shown by the dashed, dotted and full lines, respectively.
\label{fig:KK_distribution_I01}}
\end{figure}

In order to see the interferences commented above in more detail, we evaluate separately the contributions to $\bar B^0 \to J/\psi K^+ K^-$ decay from the $I=0$ and $I=1$ parts of the amplitude. This can be seen in Fig. \ref{fig:KK_distribution_I01}. What we observe there is that both the $I=0$ ($f_0(980)$) and $I=1$ ($a_0(980)$) individual contributions are larger than the total, and we also see that the shape of either of them resembles move the one of Fig. \ref{fig:mass_distribution_pi2eta} for $K^0 K^-$. The strength of the $a_0(980)$ contribution in the  $\bar B^0 \to J/\psi K^+ K^-$ decay is still smaller than that of  $B^- \to J/\psi K^0 K^-$ decay
(Fig. \ref{fig:mass_distribution_pi2eta}) because the latter contains also a tree level contribution, which is absent in the $\bar B^0 \to J/\psi K^+ K^-$ case. We also observe in Fig. \ref{fig:KK_distribution_I01}, that when we add the $I=0$ and $I=1$ parts of the amplitude, the total strength becomes smaller, indicating a strong cancellation, which is also responsible for the change in the shape of the distribution. It would be most instructive to see all these features in actual experiments.

\section{Conclusions}
  We have studied the decay rates for $\bar B^0_s \to J/\psi K^+ K^-$, $\bar B^0 \to J/\psi K^+ K^-$,
$\bar B^0 \to J/\psi \pi^0 \eta$, $B^- \to J/\psi \pi^- \eta$, $B^- \to J/\psi K^0 K^-$,
and have compared them to the rates obtained for the $\bar B^0_s \to J/\psi \pi^+ \pi^-$
 and $\bar B^0 \to J/\psi \pi^+ \pi^-$. New measurements are underway in the LHCb collaboration
 and predictions prior to the experiment are most opportune to give strength to the claims of
 the chiral unitary theory on the interpretation of resonances and the ability of the approach
 to describe the meson-meson interactions. We find that the rates of $S$-wave $K^+ K^-$ production
 are small compared to those of $\pi^+ \pi^-$ production but when integrated over the invariant mass
 they are still smaller but of the same order of magnitude. The interesting feature of the results
 is that the $K^+ K^-$ distribution peaks around the $K^+ K^-$ threshold as a consequence of
 the presence of the $f_0(980)$ and $a_0(980)$ resonances. In the case of the
$\bar B^0_s \to J/\psi K^+ K^-$, only the $f_0(980)$ resonance influences this distribution,
but in the case of the $\bar B^0 \to J/\psi K^+ K^-$, both the $f_0(980)$ and $a_0(980)$ resonances contribute to its strength.
  In the case of the  $\bar B^0 \to J/\psi \pi^0 \eta$, one finds a peak for the $a_0(980)$ resonance
and its strength is much larger than the one for  the $\bar B^0 \to J/\psi \pi^+ \pi^-$ reaction at the $f_0(980)$ peak.
We also calculated the $\pi^- \eta$ and $K^0 K^-$ mass distributions for the decays $B^- \to J/\psi \pi^- \eta$
and $B^- \to J/\psi K^0 K^-$. We found in this case that the strength of the $\pi^- \eta$ distribution is twice
the one of $\pi^0 \eta$ in $\bar B^0$ decay, and the strength of the $K^0 K^-$ distribution about four times
bigger than that of the $K^- K^+$ for the $\bar B^0$ decay. We could find an easy explanation based on the fact
that the $K^0 K^-$ production, in $I=1$, is only influenced by the $a_0(980)$ state
which has a wider distribution than the $f_0(980)$.
One interesting aspect of the calculations is that we could predict all these mass distributions with no free parameters,
up to a global normalization which is the same for all processes. The method relies on the constancy of the $V_P$ factor
which summarizes the weak amplitudes and the hadronization procedure. The only thing demanded is
that this factor is smooth and practically constant as a function of the invariant masses in the limited range
where the predictions are made. A discussion on this issue and support for this approximation
is found in Refs. \cite{liang,sekisemi}.

The predictions made here compare reasonably well with present
experimental information, but more precise data are coming from LHCb and
comparison with these data will be useful to make progress in our
understanding of the meson-meson interaction and the nature of the low
lying scalar mesons.

\section*{Acknowledgments}
We would like to thank Tim Gershon for a careful reading of the manuscript and useful information concerning the LHCb experiments.
One of us, E. O. wishes to acknowledge support from the Chinese Academy
of Science in the Program of Visiting Professorship for Senior International Scientists (Grant No. 2013T2J0012).
This work is partly supported by the Spanish Ministerio
de Economia y Competitividad and European FEDER funds
under the contract number FIS2011-28853-C02-01, FIS2011-
28853-C02-02, FIS2014-57026-REDT, FIS2014-51948-C2-
1-P, and FIS2014-51948-C2-2-P, and the Generalitat Valenciana
in the program Prometeo II-2014/068. This work is also partly supported by the
National Natural Science Foundation of China under Grant Nos. 11165005, 11565007, and 11475227.
It is also supported by the Open Project Program of State
Key Laboratory of Theoretical Physics, Institute of Theoretical
Physics, Chinese Academy of Sciences, China (No.Y5KF151CJ1).
We acknowledge
the support of the European Community-Research Infrastructure
Integrating Activity Study of Strongly Interacting Matter
(acronym HadronPhysics3, Grant Agreement n. 283286) under
the Seventh Framework Programme of EU.

\bibliographystyle{plain}

\end{document}